\documentclass [12pt,a4paper]{article}

\usepackage{amsmath}
\usepackage{amssymb}
\input epsf

\begin{document}

\thispagestyle{empty} 
%\vspace*{-80pt} 
%{\hspace*{\fill} Preprint-KUL-TF-} 
\vspace{40pt} 
\begin{center} 
{\huge\bf Quantum dynamics and \\[14pt] 
Gram's matrix} 
\vspace{40pt}

{\large M.~De~Cock, M.~Fannes\footnote[1]{Onderzoeksleider FWO Vlaanderen}, 
P.~Spincemaille} \\[7pt]   
{Instituut voor Theoretische Fysica} \\  
{Katholieke Universiteit Leuven} \\  
{Celestijnenlaan 200D} \\  
{B-3001 Heverlee, Belgium}
\end{center} 
\vspace{40pt}

\begin{abstract}
We propose to analyse the statistical properties of a sequence of
vectors using the spectrum of the associated Gram matrix. Such
sequences arise e.g.\ by the repeated action of a deterministic kicked
quantum dynamics on an initial condition or by a random process. We
argue that, when the number of time-steps, suitably scaled with respect
to $\hbar$, increases, the limiting eigenvalue distribution of the Gram
matrix  reflects the possible quantum chaoticity of the original system
as it tends to its classical limit. This idea is subsequently applied
to study the long-time properties of sequences of random vectors at the
time scale of the dimension of the Hilbert space of available states. 
\vspace{40pt}


\end{abstract}

\newpage

Discretising a classical dynamical system is in order if we want to
simulate it on a computer. Its compact phase space may for this purpose
be covered by a large number $N$ of small patches of Lebesgue measure
$1/N$. The evolution, which we assume measure preserving and discrete
in time, translates approximately in a bijection of the patches. Such a
description involves always an approximation as patches change their
shape in the course of time. It is of course helpful for actual model
systems to label the patches in a way that mimics the kinematic
structure of phase space. So, we obtain after this coarse-graining
procedure for each $N$ a one to one transformation $\pi$ of the set
$\{1, 2, \ldots, N\}$ that determines the evolution during one tick of
the clock.  The phase portrait consists in partitioning the discrete
phase space into closed orbits of $\pi$ and the crucial information is
the number of orbits together with their length as a function of $N$.
An ergodic island of non-zero measure in the dynamical system will
signal its presence by the occurrence of an orbit with a period
proportional to $N$. Iterating the dynamical map on an initial point
$i_0$ provides us with a sequence $\boldsymbol{i}= \Bigl(i_0,\, \pi(i_0),\,
\pi^2(i_0),\, \ldots \Bigr)$ of points in $\{1, 2, \ldots, N\}$ and
we can distinguish between  points belonging to ergodic or regular
regions of phase space by examining the period of the time sequence of
$i_0$ as a function of $N$.   
    
Truly quantum dynamical systems with compact phase space are finite
dimensional in virtue of the uncertainty principle. As each state
occupies a same volume $\hbar$, the dimension of their Hilbert space of
states is $1/\hbar$. Planck's constant has here a rather symbolic
meaning: for $d$-dimensional systems it is the $d$-th power of
the actual Planck constant, while $\hbar=1/(2j+1)$ for a spin with
angular momentum $j$. The $d$-dimensional complex Hilbert space, or,
more precisely, the space of complex rays in $\boldsymbol{C}^d$ is the
quantum space with $d$ elements. The space of rays is called the
projective Hilbert space of dimension $d$ and denoted by ${\rm
pr}\boldsymbol{C}^d$. In dimension 2, it turns out to be the unit sphere in
$\boldsymbol{R}^3$ with antipodal points identified. In contrast to a
classical space, distinct points can lie arbitrarily close, the
distance between the rays $[\varphi]= \boldsymbol{C} \varphi$ and $[\psi]=
\boldsymbol{C} \psi$ generated by the normalised vectors $\varphi$ and $\psi$
being
\begin{equation}
  {\rm d}\Bigl([\varphi], [\psi]\Bigr):= \inf_{z\in\boldsymbol{C},\ |z|=1}
  \|\varphi- z\,\psi\|= 2- 2|\langle \varphi, \psi \rangle|= 4 \sin^2
  \frac{\theta}{2} \ ,   
\end{equation}   
with $\theta\in[0,\pi/2]$ the angle between the rays. The maximal separation
between points is reached when they correspond to orthogonal rays. 
Projective Hilbert spaces carry a natural Riemannian structure given by
the Study-Fubini metric, but we shall not be so much concerned 
here with this continuum feature and rather focus on their discrete
aspects.

A quantum evolution in discrete time, also called kicked evolution, is 
determined by a unitary Floquet operator $u$. In Schr\"odinger picture,
$\varphi\mapsto u\,\varphi$ is the evolution between two
consecutive kicks. We use the same notation to denote the
corresponding  evolution in the space of rays: $p\mapsto u\,p$. We now
face the problem of studying time sequences $\boldsymbol{p}= \Bigl(p_0,\,
u\, p_0,\, u^2\, p_0, \ldots \Bigr)$ generated by a Floquet operator
$u$ as it acts repeatedly on an initial condition $p_0$.

In the vast literature on quantum chaos, dynamical properties of
quantum systems are often investigated by considering the temporal
behaviour of the Husimi or the Wigner functions corresponding to
well-localised states in phase space, such as coherent states. Such a
description relies, in taking the (semi)-classical limit, on a definite
geometrical picture of the corresponding classical phase space i.e.\ on
a particular choice of basic observables such as the usual position and
momentum or angular momentum. Many references can be found in~\cite{CC}
and~\cite{D}.  We argue in this letter that it is worthwhile to put
things in a more abstract perspective. Gram's matrix, or rather its
spectrum, provides us with a powerful tool to analyse the statistical
properties of time-sequences of points in a projective Hilbert space.

The Gram matrix $\mathrm{G}(\boldsymbol{\varphi})$ of a sequence
$\boldsymbol{\varphi}= \Bigl(\varphi(1), \varphi(2), \ldots,
\varphi(K)\Bigr)$ of vectors is 
\begin{equation}
  {\rm G}(\boldsymbol{\varphi})= \left( 
  \begin{array}{cccc}
  \Bigl\langle \varphi(1), \varphi(1) \Bigr\rangle 
  &\Bigl\langle \varphi(1), \varphi(2) \Bigr\rangle 
  &\ldots 
  &\Bigl\langle \varphi(1), \varphi(K) \Bigr\rangle \\
  \Bigl\langle \varphi(2), \varphi(1)\Bigr\rangle 
  &\Bigl\langle \varphi(2), \varphi(2) \Bigr\rangle
  &\ldots 
  &\Bigl\langle \varphi(2), \varphi(K) \Bigr\rangle\\
  \vdots 
  &\vdots
  &\ddots
  &\vdots \\
  \Bigl\langle \varphi(K), \varphi(1) \Bigr\rangle 
  &\Bigl\langle \varphi(K), \varphi(2) \Bigr\rangle
  &\ldots 
  &\Bigl\langle \varphi(K), \varphi(K) \Bigr\rangle
  \end{array}
  \right) \ .
\end{equation}  
${\rm G}(\boldsymbol{\varphi})$ is positive definite and its rank
equals the dimension of the space spanned by the
$\varphi(j)$~\cite{HJ}. In particular $\boldsymbol{\varphi}$ is
linearly independent if and only if $\det({\rm
G}(\boldsymbol{\varphi}))\ne 0$. The spectrum of ${\rm
G}(\boldsymbol{\varphi})$ is independent on the order of the
$\varphi(j)$ in $\boldsymbol{\varphi}$ and on multiplying the
$\varphi(j)$ with a complex number of modulus 1. This means that for a
given sequence $\boldsymbol{p}= \Bigl(p(1),\, p(2),\, \ldots,\,
p(K)\Bigr)$ of points in a projective Hilbert space, specified by
normalised vectors $\varphi(j)$ as $p(j)= [\varphi(j)]$,  the spectrum
of ${\rm G}(\boldsymbol{\varphi})$ depends only on $\boldsymbol{p}$ and
that it is insensitive to the order of the points in $\boldsymbol{p}$.
It may therefore be denoted by $\Sigma(\boldsymbol{p})$. 

Let us for a moment consider a classical word $\boldsymbol{i}= 
\Bigl(i(1),\, i(2),\, \ldots,\, i(K)\Bigr)$ where the letters are
chosen from a given alphabet $\{1,2,\ldots\}$. In fact,
$\boldsymbol{i}$ is  in one to one correspondence with a sequence
$\boldsymbol{p}= \Bigl([e_{i(1)}],\, [e_{i(2)}],\, \ldots,\,
[e_{i(K)}]\Bigr)$ of vectors  through the identification of $j$ with
$e_j$ for an orthonormal basis $\{e_1,\, e_2,\, \ldots\}$ of a Hilbert
space. Grouping $e_{i(\ell)}$ with equal index $j$, the Gram matrix is
block-diagonal with block $E(j)$ of the type
\begin{equation}
  \left( 
  \begin{array}{cccc}
  1 &1 &\ldots &1 \\
%  1 &1 &\ldots &1\\
  \vdots &\vdots &\ddots &\vdots \\
  1 &1 &\ldots&1
  \end{array}
  \right) \ .
\end{equation} 
The dimension of $E(j)$ is precisely the multiplicity $m(j)$ of $j$ in
$\boldsymbol{i}$. As the spectrum of $E(j)$ consists of the
non-degenerate eigenvalue $m(j)$ and the $m(j)-1$ degenerated
eigenvalue 0, we find that $\Sigma(\boldsymbol{p})$ determines
precisely the amount of different numbers appearing in $\boldsymbol{i}$
with their multiplicity i.e.\ the relative frequencies of the different
letters in $\boldsymbol{i}$. The spectrum of the Gram matrix of a very
regular  sequence $\boldsymbol{i}$ will consist of a few large naturals
and a highly degenerated 0 while for sequences with many different
indices the spectrum will be concentrated on small natural numbers
appearing with high multiplicities. A same interpretation remains valid
for the general non-commutative case: a Gram matrix with spectrum
concentrated around small natural numbers points at a vector wandering
wildly through the Hilbert space of the system and is therefore a sign
of chaotic behaviour. More regular motion, such as precession or slow
diffusion, signals its presence by large eigenvalues and a high
occurrence of eigenvalues close to 0. In contrast to the classical case
however, eigenvalues are no longer limited to natural numbers so that a
same point in a projective Hilbert space can now be visited a
fractional number of times. 

The existence for quantum systems of an intermediate time scale with
interesting and describable behaviour in between the very short one, of
order $-\log \hbar$, where the quantum system slavishly follows its
classical limit and the very long one where quasi-periodic behaviour,
due to the discreteness of the spectrum of the Floquet operator, is
dominant, is a central theme in many papers~\cite{CC2}. We are, more
precisely, interested in the limiting eigenvalue distribution of the
Gram matrix when its dimension, the number of time-steps, tends to
infinity, appropriately scaled with respect to the quantum parameter.
To obtain this, we consider the limit of the empirical measure
\begin{equation*}
  \frac{1}{K} \sum_{j=1}^K \delta(\lambda- \gamma_j)
\end{equation*}
where the $\gamma_j$ are the eigenvalues of the Gram matrix. The
limiting distribution should reflect the quantum character of the
dynamics as it tends to its classical limit. We do not claim that we
can settle this question but we shall at least provide some rigorous
support for this expectation. 

Instead of considering a genuine unitary dynamics acting on an initial
condition, we consider sequences $\boldsymbol{p}= \Bigl( p(1),\, 
p(2),\, \ldots,\, p(K)\Bigr)$ of points in the projective Hilbert space
of dimension $N$, independently and randomly chosen with respect to the
uniform measure. Recall that ${\rm pr}\boldsymbol{C}^N$ is compact and
that it carries a unique normalised measure, called uniform, which is
invariant under the action of every $N\times N$ unitary. Picking
independent normalised vectors randomly with respect to this measure is
quite different from picking the components of the vectors with respect
to a given basis in an independent and random way with respect to some
suitably chosen probability measure. Next, we compute the spectrum of
the Gram matrix of such a sequence. This spectrum is of course a random
object but it turns out that, in the limit of large $N$ and for a
rescaled time $\tau= K/N$, the spectral distribution tends  to a
definite limit given by the Marchenko-Pastur distribution
$\mu_\tau$~\cite{MP}. The actual computations are somewhat involved and
will be presented in~\cite{DFS}. Though the Gram matrices are given in
terms of independent random vectors, there is no independence between
the entries. E.g., each Gram matrix is positive definite which is
totally incompatible with independence of matrix elements. This makes
the computation quite different from Wigner's random matrix
computation. $\mu_\tau$ is obtained either by a combinatorial argument
in terms of its moments or by determining the expectation of the
resolvent of the Gram matrix. 
     
When $0< \tau\le 1$ the probability measure $\mu_\tau$ is given by a
continuous density $\rho$. For very small $\tau$, we must choose
relatively few vectors in a large space. This will often lead to almost
orthogonal choices and therefore $\rho$ will be concentrated around 1.
When $\tau$ increases to 1, there is a fair chance of many vectors
overlapping and the support of $\rho$ will simultaneously extend
towards 0, which is a lower bound of its support, and to larger
positive values. When $\tau>1$ there will almost surely be a sizeable
degree of linear dependence responsible for a high multiplicity of the
eigenvalue 0. In fact, it turns out that $\mu_\tau$ decomposes for
$\tau>1$ into an atom at 0 and an absolutely continuous part:
\begin{equation}
  d\mu_\tau(x)= \frac{\tau-1}{\tau}\, \delta(x)\, dx+ 
  \rho(x)\, dx \ ,  
\end{equation}  
where $\rho$ is the absolutely continuous part of $\mu_\tau$. The
weight of $\rho$ is $1/ \tau$. Moreover $\rho$ is compactly supported
in the interval $[(\sqrt\tau-1)^2, (\sqrt\tau+1)^2]$. A similar
computation in the classical case yields a Poisson distribution. This
result is reminiscent of Wigner's semicircular distribution for the
spectrum of large random matrices where compactness of the limiting
distribution is also a typical feature of the
non-commutativity~\cite{M}. A simple measure of the dynamical entropy
of the system is the length of the support of $\mu_\tau$. For our
random dynamics, this quantity grows as $4 \sqrt{\tau}$, in contrast to
an expanding chaotic dynamics where the entropy grows linearly in
$\tau$.

\hspace{2cm}
\begin{center}
  \epsfxsize=15cm \epsfbox{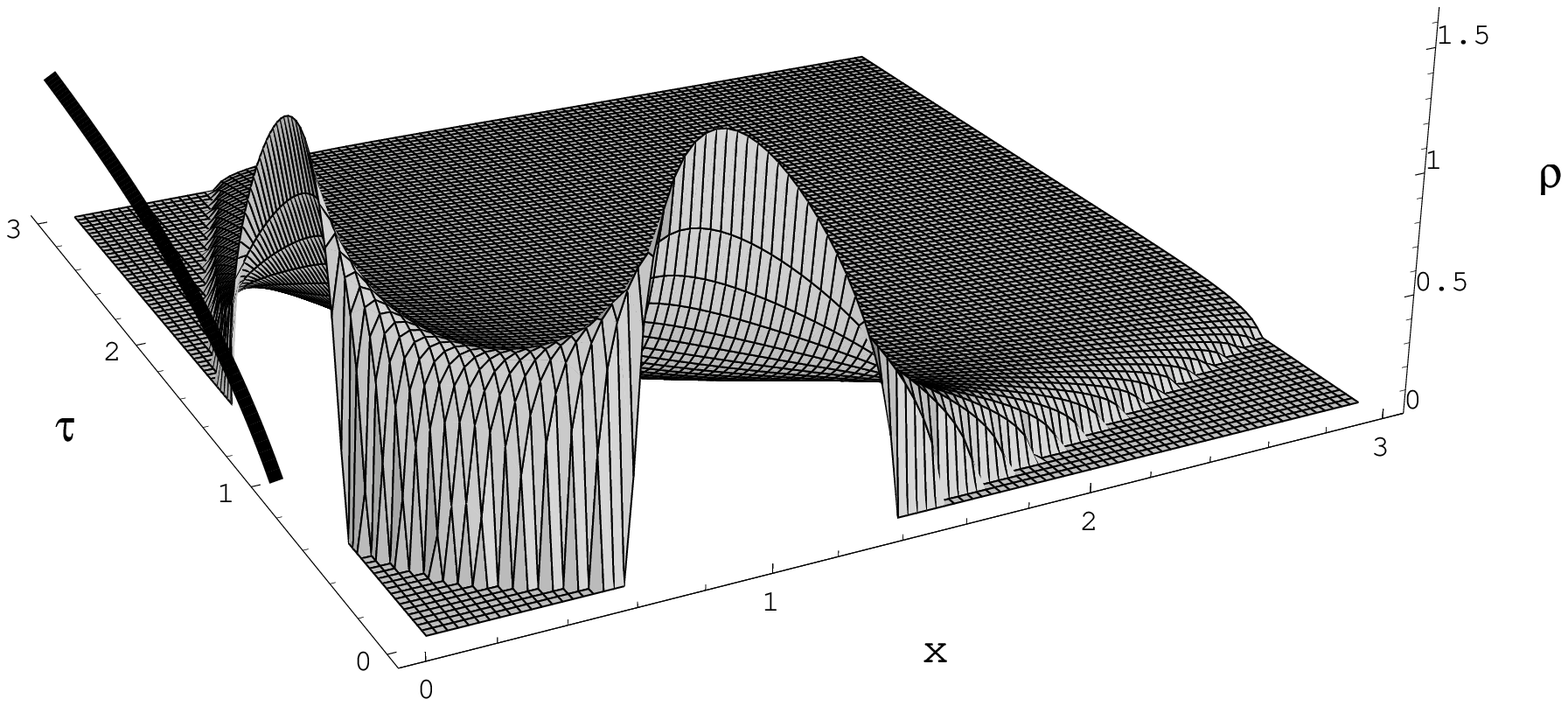}
{\bf Figure 1}: The limiting spectral distribution of $G(\boldsymbol{p})$
\end{center}
\hspace{2cm}

The figure shows the limiting spectral distribution of Gram matrices in
the region $0.02\le \tau\le 3$. The $\delta$ contribution of weight
$(\tau-1)/\tau$ that appears for $\tau>1$ is rendered by the fat line,
which has height $(\tau-1)/\tau$. The continuous part is, for all
values of $\tau$, only non-vanishing  for $(\sqrt\tau-1)^2< x<
(\sqrt\tau+1)^2$, but this is only visible in the figure for moderately
small values of $\tau$. For $\tau$ tending to 0, a $\delta$
distribution at $x=1$ will appear and the probability density has for
$\tau=1$ a singularity at $x=0$ of order $-1/2$. 

It is a pleasure to thank H.~Wagner for his constant interest and
enthusiasm in pointing out the relevance of geometrical cum statistical 
ideas in physics. Two of the authors (M.D.C. and P.S.) acknowledge financial
support from FWO project G.0239.96.

\noindent
{\Large\bf Figure Captions}
\bigskip

\noindent
{\bf Figure 1}: The limiting spectral distribution of $G(\boldsymbol{p})$

\end{document}